\documentclass[twocolumn,showpacs,amsmath,amssymb,longbibliography,superscriptaddress,showkeys]{revtex4-1}
\usepackage{color,longtable}
\usepackage{graphicx}
\usepackage{amssymb,amsfonts,amsmath}
\usepackage{here}
\begin{document}
\title{Spreading of foam on a substrate}
\author{Masaya Endo, Marie Tani and Rei Kurita}

\affiliation{%
Department of Physics, Tokyo Metropolitan University, 1-1 Minamioosawa, Hachiouji-shi, Tokyo 192-0397, Japan
}%
\date{\today}
\begin{abstract}
Foam is an industrially important form of matter, commonly deployed to clean objects and even our own skin, thanks to its ability to absorb oil and particles into its interior. To clean a large area, a foam is spread over a substrate, but the optimum conditions and mechanism have been unclear. Here, we study how a foam is spread by a rigid plate on a substrate as a function of spreading velocity, gap height, confinement length, amount of foam and wettability of the substrate. Three distinguishable spreading patterns were found: homogeneous spreading, non-spreading, and slender spreading. 
It is also found that the dynamics and the mechanism of the spreading can be explained by coupling among dewetting, anchoring, shear stress, viscous stress and yield stress. 
It is a unique feature of foams, which is not observed in simple liquids and then
these findings are also critical for understanding the mechanical response of other soft jamming systems such as cells and emulsions.
\end{abstract}

\maketitle
A foam is composed of a collection of soap bubbles, but it is not simply a sum of its parts; foams have a variety of unique properties that distinguish them from individual bubbles~\cite{weaire2001, cantat2013}. For example, foams have the ability to absorb oil and tiny particles into its interior. They can also block heat and mass transport from one side to the other, being composed almost entirely of air. And though they are composed of fluid, they also possess elastic properties. These properties are utilized in everyday applications such as in foods, detergents, cosmetics, and fire extinguishing agents. Foams can also trap microorganisms, providing a facile route to improving water quality ~\cite{Roveillo2020}.

The mechanisms by which foam properties are expressed are highly complex, since many physicochemical factors play a role; examples include the mean and distribution of bubble sizes, surface tension, and interfacial rigidity of the surfactant ~\cite{Biance2011, Bera2013, Langevin2017, Furuta2016, Kurita2017}. Key among these is the ability for soft bubbles in a liquid to jam~\cite{Merrer2012, Katgert2013, Yanagisawa2021a}. This is manifested in many ways. Foams have a finite elasticity in response to shear stress in the zero-frequency limit; the viscosity of foams decreases at high shear rates, a phenomenon known as shear thinning~\cite{cantat2013, kraynik1988, cohen2013, hohler2005}; furthermore, bubbles inside a foam are collectively rearranged by small perturbations ~\cite{Katgert2013, Yanagisawa2021a,Yanagisawa2023}. These dynamical features may come from the features of jammed systems. Thus, the dynamics of a foam are not the same as that of a simple liquid, yet such properties remain to be fully understood.

Here, we focus on how a foam is spread by a rigid plate. 
Historically, the spreading of a simple liquid by a rigid plate is a well-studied problem~\cite{taylor1962}. The thickness of a spread liquid film $e$ depends on the spreading velocity $V$, following $e \propto V^{2/3}$~\cite{levich1942, derjaguin1941,bretherton1961}. Spreading using an elastic scrapper has also been reported for a simple liquid~\cite{seiwert2013,krapez2020}, and a complex fluid ~\cite{krapez2022}, whose thickness follows a modulated scaling law. 
Note that a foam is a jammed system with a tendency to deform macroscopically, since the bubbles are frictionless and deformable. Thus, it is expected that the dynamics of a spreading foam should be different from that of a simple liquid. 
While spreading of the foam is a ubiquitous process in everyday life; we often spread hand soap, dish washing liquid, shaving foam, etc. over large surfaces for cleaning, lubrication, protection, etc., no experimental studies focusing on the foam spreading process has been reported. More generally, the behavior of foam when a shear stress is applied to it on a substrate has not been clarified.

The purpose of this study is to understand how foams are spread by a rigid plate while changing the spreading velocity, the gap between the plate and the substrate, the confinement length, the amount of foam and the wettability of the substrate. The results of this study are industrially important; understanding spreading behavior and its mechanism may lead to more effective applications. This study also provides critically important information for understanding the kinetics of soft jamming such as in emulsions and cells. 

We used a 5.0 wt\% solution of ionic surfactant TTAB (tetradecyl trimethyl ammonium bromide) in glycerol and deionized water. This TTAB concentration is higher than the critical micelle concentration i.e. 0.12 wt\%~\cite{Danov2014}, and the surface tension at 3.0 wt\% is 37 mN/m. It is known that the interface rigidity of TTAB is quite small~\cite{Yanagisawa2023}. In order to prevent the collapse of foams, we used 40 wt\% glycerol. The density of the solution $\rho$ is 1.10 g/cm$^3$, and the viscosity $\eta$ of 40 wt\% glycerol solution is 4.7×10$^{-3}$ Pa$\cdot$s, respectively. We used a pump to produce foams with a liquid fraction of $\phi = 0.18$. The mean bubble size $d$ is 0.28 mm and the standard deviation is 0.08 mm. The radius of curvature $r$ of Plateau borders is about 0.1 mm. We also used a 20 wt\% dilution of a commercial dish soap (Charmy, Lion Co., Japan), and confirmed that the results are essentially same as in the TTAB system (as described in the Supplementary Information~\cite{Supple}). 

\begin{figure}[t]
\begin{center}
\includegraphics[width=65mm]{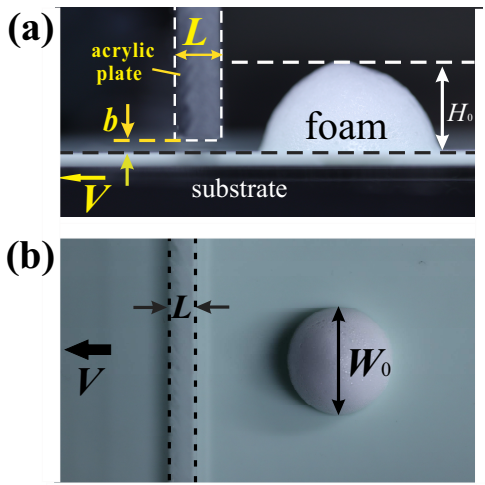}
\end{center}
\caption{Experimental setup (a) from the side and (b) from the top. A foam is put on the substrate, and the substrate is moved horizontally at a constant velocity $V$ using an electric slider. The foam is spread by a fixed acrylic plate with confinement length $L$. The gap height between the substrate and the plate $b$ and the diameter of the foam $W_0$ can be controlled. We recorded the spreading dynamics from the top using a video camera. }
\label{setup}
\end{figure}

Our experimental setup is shown in Fig.~\ref{setup}. We put a foam on a substrate and set an acrylic plate in front of it. The gap between the acrylic plate and the substrate $b = 1.0 - 2.5$ mm and the confinement length $L = 1 - 10$ mm were controlled independently. The initial shape of the foam was almost a hemisphere, with diameter $W_0$ and height $H_0$ of about 20 mm and 12 mm when $\Omega$ = 3.0$\times$10$^3$ mm$^3$. The initial diameter $W_0$ was varied from 15 to 30 mm by varying the amount of foam. As a substrate, we mainly used a hard elastic sheet made of polydimethyl-siloxane (Correcsil puls, Yamahachi Dental MFG., Co., Japan) to obtain a partially wetting surface. The static contact angle of this substrate with the solution is $\theta_E$ = 46 $\pm$ 2$^\circ$. We call this substrate the ``low wettability substrate". We also used an acrylic substrate with $\theta_E$ = 23 $\pm$ 2$^\circ$. We refer to this acrylic substrate as the ``high wettability substrate". To spread the foam, the substrate was moved horizontally at a constant speed $V = 1.0 - 35.0$ mm/s using an electric slider (LTS300/M, Thorlabs, Inc., US). The foam is spread by the fixed acrylic plate with a confinement length $L$. The spreading dynamics were recorded from above using a video camera (EOS R, Canon Inc., Japan).

\begin{figure*}[ht]
\begin{center}
\includegraphics[width=180mm]{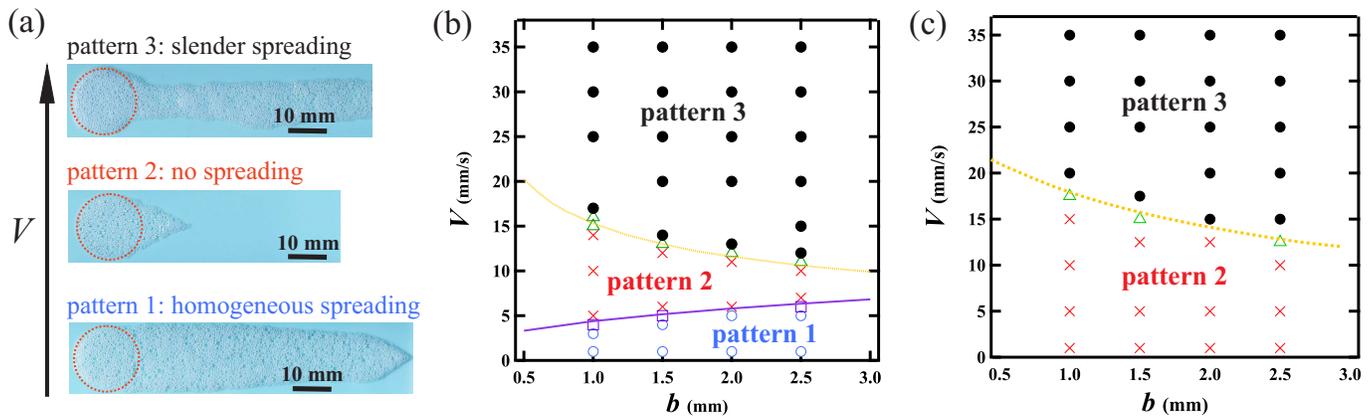}
\end{center}
\caption{(a) Typical images of spreading observed on a low wettability substrate: homogeneous spreading (pattern 1) where the foam is homogeneously spread ($V$ = 2.0 mm/s), non-spreading (pattern 2) where the foam mostly slips on the substrate and some liquid remains ($V$ = 10.0 mm/s), and slender spreading (pattern 3) where the foam is spread along a neck-like profile ($V$ = 25.0 mm/s), respectively. Other parameters are kept the same: $b = 1.5$ mm, $L = 5.0$ mm and the amount of the foam $\Omega \simeq 3.0 \times$ 10$^3$mm$^3$ i.e. $W_0 = 20$ mm, respectively. The dotted circles represent the initial positions of the foams. Diagram of the observed patterns as a function of $V$ and $b$ on (b) the low wettability and (c) high wettability substrates. Other parameters are kept the same: $L = 5.0$ mm and $W_0 = 20$ mm, respectively. The solid curve i.e. the boundary between patterns 1 and 2 indicates Eq.~(\ref{eq:Vc1}) with a fitting parameter $V^*$ = 0.47 mm/s. The dashed curve shows the boundary $V_{c2}$ between patterns 2 and 3 and is as a guide for the eye.
}
\label{phase}
\end{figure*} 

Firstly, we investigate the spreading behavior of the foam both on low and high wettability substrates. We note that foam spreading on low wettability substrates is ubiquitous in our daily lives. Examples include foams on an oil layer, on hydrophobic skin, or on a hydrophobic coating. We found three distinguishable spreading patterns on a low wettability substrate, as shown in Fig.~\ref{phase}(a): homogeneous spreading (pattern 1), non-spreading (pattern 2) and slender spreading (pattern 3) (Movies in SI~\cite{Supple}). The observed pattern depends on the spreading velocity $V$ and the gap $b$, as shown in Fig.~\ref{phase}(b). 
On low wettability substrates, bubbles which are initially in contact with the substrate are fixed at their initial positions, irrespective of the pattern. 
Meanwhile, it is found that this is not the case for high wettability substrates as shown in Fig.~S1~\cite{Supple} and then pattern 1 cannot be observed on a high wettability substrate.
We note here that those behaviors cannot be observed in simple liquid systems.

Here, we show the spreading dynamics when each pattern is formed. When $V$ is small, the foam is homogeneously spread (``pattern1: homogeneous spreading''; open circle symbols in the diagram). 
The base of the foam can be anchored on a low wettability substrate as we mentioned above. 
This anchoring is clearly an important factor for the formation of pattern, since we do not observe pattern 1 on the high wettability substrate. 
The anchored bubbles moves with the substrate and then the strong shear stress is applied to the foam. 
This is consistent with a previous report where bubbles move with the substrate when they are anchored by hydrophobic grains~\cite{marchand2020}.
Since the shear stress excesses the yield stress of the foam, the foam is subsequently flattened and spread by the plate. 
We will discuss later why the foam is anchored by the low wettability substrate.

When $V$ is in the mid-range, the foam slips without anchoring except for the bubbles initially in contact with the substrate (``pattern 2: no spreading''; cross symbols in the diagrams). During slipping, the foam leaves a footprint of liquid film on the substrate. 
Here, we estimated the thickness of the spread liquid film $e_{\ell2}$ by measuring the weight of the remaining solution $m_\ell$ as $e_{\ell2} = m_\ell/(\rho \xi W_0)$ where $\rho$, $\xi$, and $W_0$ are the density of the solution, the spreading distance, and the width of the initial foam, respectively. This gives $e_{\ell2}$ = 8.4 $\pm$ 1.5 $\mu$m, 8.8 $\pm$ 1.4 $\mu$m and 7.8 $\pm$ 0.6 $\mu$m for $b$ = 1.0, 1.5 and 2.0 mm, respectively, and $V$ = 8 mm/s, where the mean and the deviation are calculated from 5 independent measurements. Thus, $e_{\ell2}$ is independent of the gap height $b$. 
Here, the bubbles inside the foam are immobile, while the liquid is pulled out from the outermost part of the foam, forming a liquid film. This situation is similar to what happens during dip coating. 
Thus, we estimate the thickness of the liquid film using the Landau-Levich-Derjaguin law; $e_{\ell2} = \kappa^{-1} (\eta V/\gamma)^{2/3}$, where $\kappa^{-1}$ is the capillary length~\cite{levich1942, derjaguin1941}. Since $\kappa^{-1}$ = 1.9 mm for our solution, we obtain $e_{\ell2}$ = 19 $\mu$m. This value is of the same order as our experimental estimate, although it is only a rough approximation.

\begin{figure}[h!]
\begin{center}
\includegraphics[width=85mm]{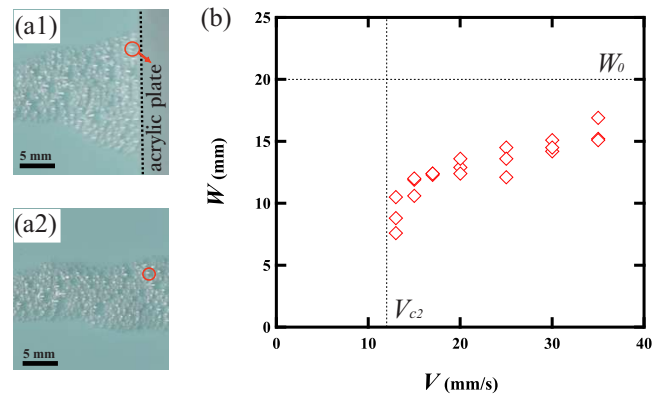}
\end{center}
\caption{(a) Images of the spread foam at $V$ = 15.0 mm/s and $b$ = 1.5 mm (pattern 3) (a1) during spreading and (a2) in the final state. Time difference between the pictures is 1 s. The circles in (a1) and (a2) represent the same bubble, which indicates that the bubbles move to the center during spreading. (b) $V$ dependence of the width $W$ of the final state of the spread foam when $b$ = 1.5 mm. $V_{c2}$ is the onset velocity of pattern 3 and $W_0$ is the initial width of the foam.
}
\label{region3}
\end{figure}

When $V$ is large, the foam is spread again (``pattern 3: slender spreading''; filled circle symbols in the diagrams). However, there are features which distinguish it from pattern 1; the width of the spread foam narrows as shown in Fig.~\ref{phase}(a). 
It is also found that some bubbles move after spreading, as shown in Fig.~\ref{region3}(a1) and (a2). 
The circle in Fig.~\ref{region3}(a1) and (a2) represents the same bubble just after spreading and at the final state, respectively. 
It is clear that the bubbles are not anchored on the substrate just after spreading.
This is also different from the behavior in pattern 1, where the bubbles cannot move after spreading.
We measured the $V$ dependence of the width $W$ of the spread foam at steady state. Figure \ref{region3}(b) shows the $V$ dependence when $b$ = 1.5 mm. $V_{c2}$ is a transition velocity between the incidence of patterns 2 and 3, and $W_0$ is the initial width of the foam before spreading. $W$ gently increases with increasing $V$ and seems to asymptotically approach $W_0$. It is reflected in the fact that the foam can be spread with longer length near $V_{c2}$.
Here, we propose a possible mechanism for the formation of pattern 3. 
The foam is pulled by the substrate via a liquid film at the base of the foam. 
When the viscous stress at the base liquid film is larger than the yield stress of the foam or $V > V_{c2}$, the foam may undergo plastic deformation and some bubbles can escape confinement (pattern 3), while the foam cannot be deformed plastically and the foam slips for $V < V_{c2}$  (pattern 2).
Thus, for larger $V$ ($> V_{c2}$), the larger amount of the bubbles escape the confinement and $W$ becomes larger. 
Other phenomena such as the collective motion of bubbles~\cite{Yanagisawa2021a} might play a role, but a more quantitative argument will require more precise measurements; this is beyond the scope of this paper.

\begin{figure}
\begin{center}
\includegraphics[width=85 mm]{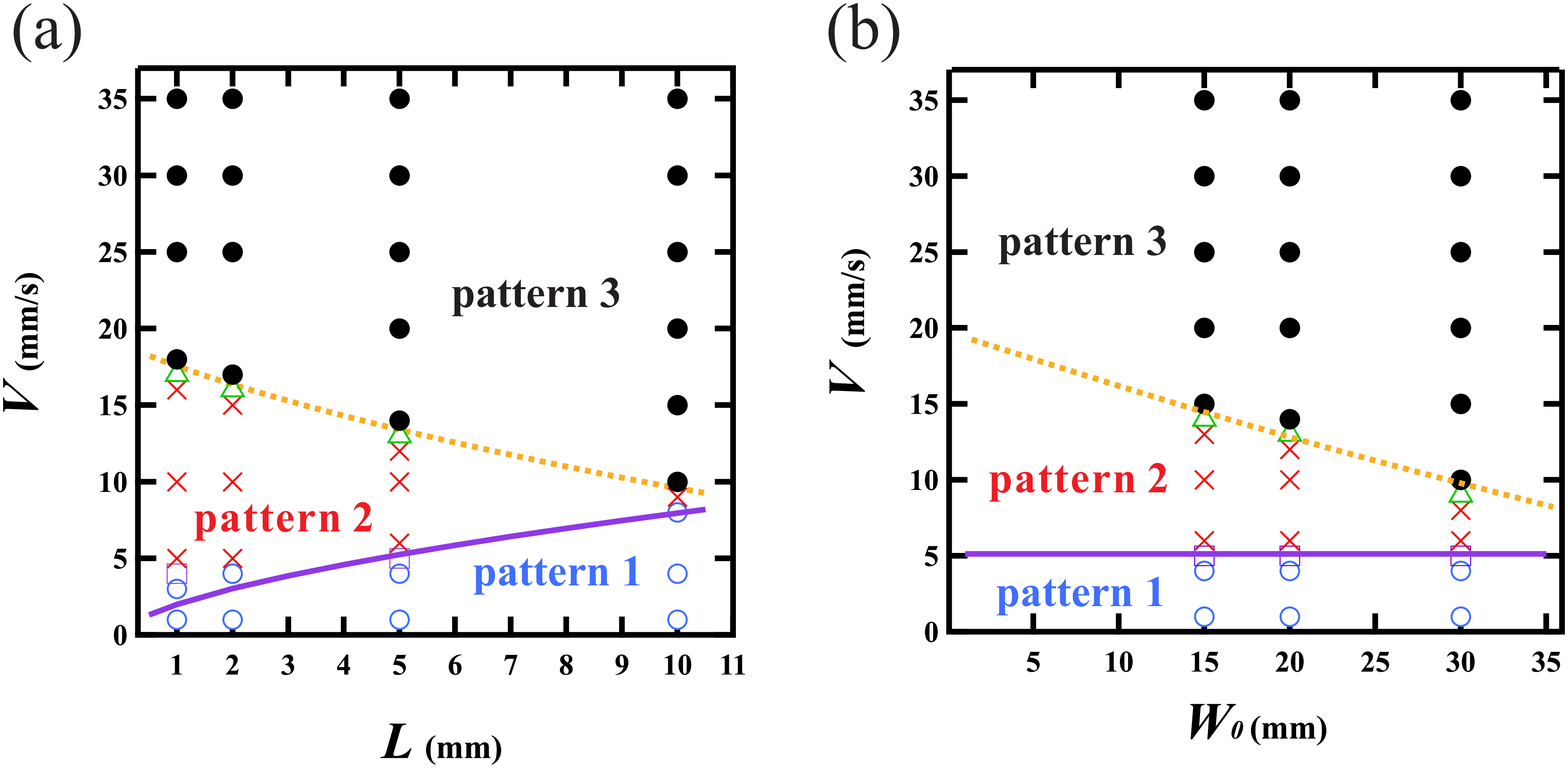}
\end{center}
\caption{
Diagrams of the observed spread patterns on the low wettability substrate as a function of $V$ and (a) $L$ ($b=1.5$ mm $W_0=20$ mm), and (b) $W_0$ ($b=1.5$ mm $L=5.0$ mm), respectively. All solid curves indicate Eq.~(\ref{eq:Vc1}) with $V^*$ = 0.47 mm/s. The dashed lines show the boundary $V_{c2}$ between patterns 2 and 3 and are a guide for the eye.
}
\label{phase2}
\end{figure}

Now, we shall describe why the foam is anchored in pattern 1 on the low wettability substrate. 
It is found that the liquid dewets the substrate after the foam slips with making a liquid film in the pattern 2.
Thus, it is expected that the anchoring of the bubbles on the substrate is caused by dewetting; more specifically, the thin film of liquid at the base of the foam dewets, leaving only Plateau borders in contact with the substrate which anchor the bubbles.
In order to estimate the dewetting velocity, we assume that our system is in a similar situation to when a liquid is confined between a rigid plate and an elastic material~\cite{brochard1994,martin1998}; this is reasonable since foams are elastic. In such a case, the initial dewetting velocity $V_d$ of the liquid layer on the substrate can be estimated as $V_d =K |S| h_0/(\eta e_{\ell1})$. Here, $K$ is a numerical prefactor and $S$ is the spreading constant i.e. $S= \gamma_S -(\gamma_{SL}+\gamma)$. $\gamma_S$ and $\gamma_{SL}$ are the interfacial tensions of solid-air and solid-liquid interfaces, written as $S=\gamma(\cos{\theta_E}-1)$ with static contact angle $\theta_E$ using Young's law~\cite{young1805}. $h_0=|S|/E$, where $E$ is the Young's modulus of the elastic material, and $e_{\ell1}$ is the thickness of the liquid film. 
We estimate $e_{\ell1}$ using $e_{\ell1} \sim r (\eta \Delta V/\gamma)^{2/3}$~\cite{bretherton1961} with $\Delta V \sim Vd/b$ where $b/d$ is the number of layers of bubbles between the plate and the substrate. 
We note that the radius of curvature of the Plateau border $r$ is taken as the typical length scale since the liquid film is formed at the base of the foam, drawing liquid from Plateau borders in the foam interior. This is different from when we estimated $e_{\ell2}$, where the liquid is spread from the outer edge. As a result, the time $\tau_d$ that a bubble of diameter $d$ needs to be in contact with the substrate can be estimated as $\tau_d \sim d/V_d$. On the other hand, the time $\tau_{shear}$ over which the foam is sheared in the confined space can be written as $\tau_{shear} \sim L/V$. Thus, pattern 1 seems to occur when $\tau_d < \tau_{shear}$ i.e., $V <V_{c1}$ where 
\begin{equation}
V_{c1} = (V^*/d) b^{2/5} L^{3/5}. 
\label{eq:Vc1}
\end{equation}
Here, $V^*= K^{3/5} (\gamma/\eta)(|\cos{\theta_E}-1|h_0/r)^{3/5}$. Equation~(\ref{eq:Vc1}) successfully describes the border between patterns 1 and 2 in the diagram with a fitting parameter $V^* = 0.47$ mm/s, as shown in Fig.~\ref{phase}(b). To examine the $L$ dependency and independence of $W_0$, we varied $L$ and $W_0$ on a low wettability substrate, and made similar diagrams as shown in Fig.~\ref{phase2}. As indicated by the solid curves in Fig.~\ref{phase2}, Eq.~(\ref{eq:Vc1}) successfully describes the borders with the same value of $V^*$. We therefore conclude that the foam is anchored because of dewetting of the substrate in pattern 1. 

In summary, we investigated the behavior of foams spread by a plate. On a low wettability substrate, it was found that the spreading foams formed three different patterns depending on spreading speed, and that the boundary between patterns varied depending on the gap between the spreading plate and the substrate, the confinement length, and the initial width of the foam. When the foam is spread slowly, the foam is anchored to the substrate, leading to a homogeneous and flat spread. At higher speeds, the foam slips since wetting due to the spreading overcomes dewetting from the substrate, forming a thin liquid film whose thickness is given by the Landau-Levich-Derjaguin's law. At even higher speeds, the foam is seen to spread again. This may occur when the viscous stress overcomes the yield stress of the foam. 
Here we note that those behaviors cannot be observed in simple liquids and it suggests that they may be typical for soft jamming systems.
Thus, our results may not only lead to more effective application of foams, but also a deeper understanding of the kinetics of soft jamming, such as in emulsions and cells. 

M. T. was supported by JSPS KAKENHI (20K14431) and R. K. was supported by JSPS KAKENHI (17H02945 and 20H01874). 

RK conceived the project. ME performed the experiments. MT and RK considered the mechanism behind the spreading behavior. All authors wrote the manuscript.


Correspondence and requests for materials should be addressed to MT (mtani@tmu.ac.jp) and RK (kurita@tmu.ac.jp).


\clearpage
\end{document}